\documentclass[final,1p,times,twocolumn,authoryear]{elsarticle}
\usepackage{amssymb}
\usepackage{lineno}
\usepackage{siunitx}
\usepackage{subcaption}

\journal{Arxiv}

\begin{document}

\begin{frontmatter}

\title{Ground water retention correlation to atmospheric muon rates}

\author[affil]{T. Avgitas}
\ead{avgitas@ip2i.in2p3.fr}

\author[affil]{J.-C. Ianigro}
\ead{ianigro@ip2i.in2p3.fr}

\author[affil]{J. Marteau}
\ead{marteau@ip2i.in2p3.fr}

\affiliation[affil]{organization={Univ Lyon, Univ Claude Bernard Lyon 1, CNRS/IN2P3, IP2I Lyon, UMR 5822},
            addressline={},
            postcode={F-69622},
            city={Villeurbanne},
            country={France}}

\begin{abstract}
Muography is an investigation technique based on the detection of the atmospheric muon flux' modification through matter. It has found lately multiple applications in geosciences, archaelogy, and non invasive industrial controls. Mostly known for its imaging capabilities, muography may be exploited as well for monitoring purposes since the atmospheric muon flux is available permanently. In this paper we present an interesting measurement performed in the context of an archaelogical project called Arch\'emuons, on the archaeological site of "Palais du Miroir" in Vienne, South of Lyon, France. We installed a muon detector in an underground gallery within the foundations of the building for the second half of 2023. The primary goal is to measure details of those foundations which are largely not excavated yet. Meanwhile we observed over more than 6 months long-term and short-term variations of the muon rates since the start of the experiment, which seem to exhibit a correlation with the rain accumulating on the free field just above the gallery. We propose as an explanation for this behavior the retention of water by the soil above the detector site. 
\end{abstract}




\end{frontmatter}


\section{Introduction}
\label{introduction}
Muon imaging or muography has emerged as a powerful non-invasive method to complement standard tools in Earth Sciences and is nowadays applied to a growing number of fields such as industrial controls, homeland security, civil engineering. This technique relies on the detection of modifications - absorption or scattering - in the atmospheric muon flux when these particles cross a target. Atmospheric muons are secondary products of primary cosmic-rays, essentially protons and helium nuclei expelled by stars, interacting with nuclei encountered on the top of the atmosphere. 

The rather low interaction cross-section of muons with matter ensures that most of them reach the Earth’s ground level and that furthermore they may significantly penetrate large and dense structures. As suggested originally by Alvarez in 1970 for the Chephren pyramid \citep{alvarez1970}, this property may be exploited to perform density contrasts analysis of the interior of the target like X-rays do in medical imaging. As suggested by previous works in volcanology \citep{jourde2016}, geology \citep{tramontini2019} or atmosphere surveys \citep{direnzo2021}, the permanence of the muons flux may find applications in monitoring the changes in the inner part of targets under study. This is one of the goal of the Archémuons project running since 2023 in Vienne, South of Lyon, France. The primary objective of the project is the characterization of underground galleries in a barely excavated archaeological site. To this end Archémuons collaboration consists of three institutes, IP2I of Lyon, Archéorient\footnote{French National Centre for Scientific Research CNRS, UMR 5133, Arch\'eorient, Lyon, France} and LGL-TPE\footnote{Laboratoire de Géologie de Lyon: Terre, Planètes, Environnement, Université de Lyon, Université Lyon 1 and Ecole Normale Supérieure de Lyon, UMR CNRS 5276, F-69622 Villeurbanne, France} with the latter two being in charge of the geophysics surveys related to the properties of the soil that covers the galleries.

\begin{figure}[h]
    \centering
    \includegraphics[width=0.80\textwidth]{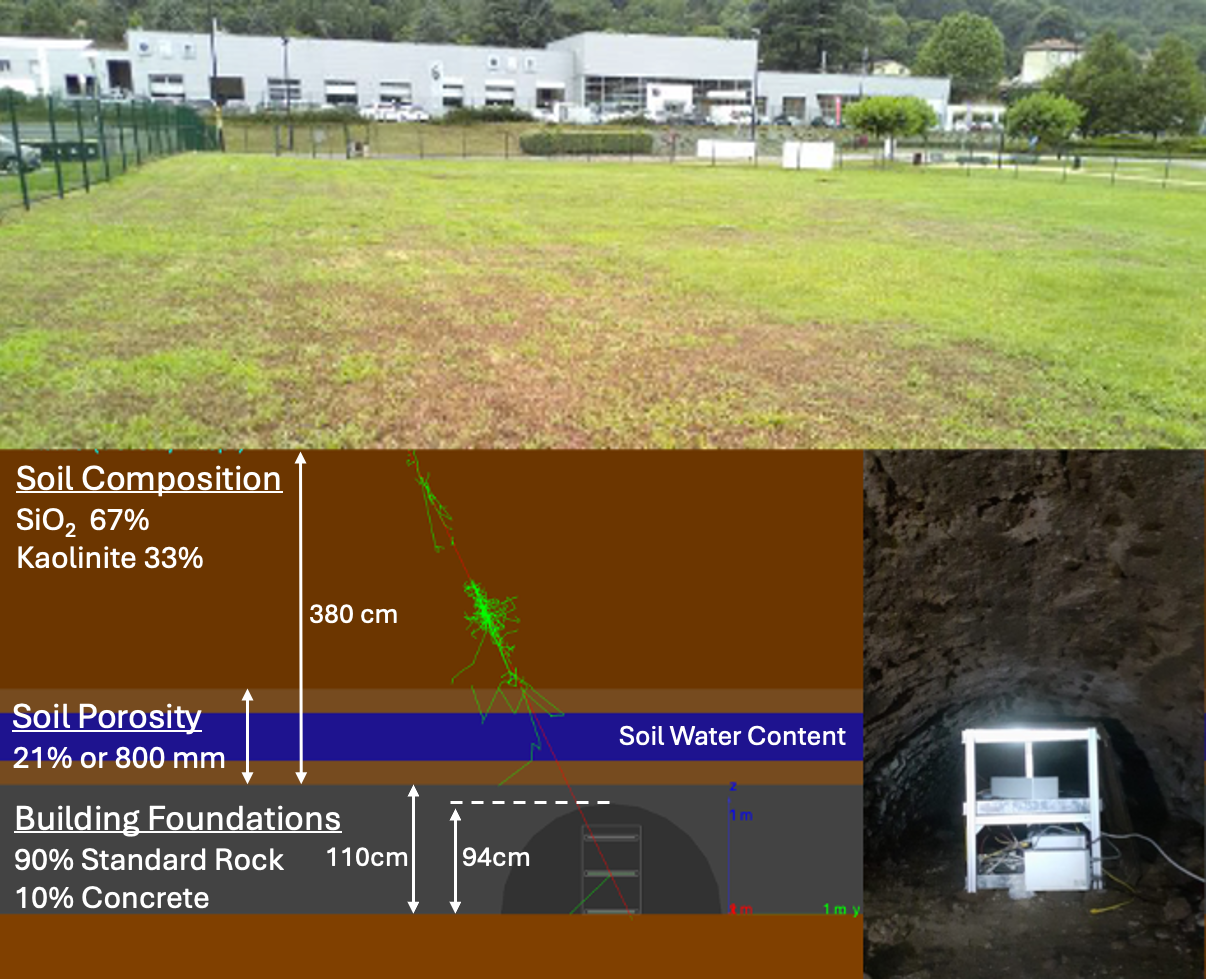}
    \caption{ top - The field above the foundations of the archaeological site of Palais du Miroir. Bottom (left) - The simulated version of our experimental setup, the details are discussed inside the text. Bottom (right) - Photograph of the muon tracker installed inside the foundations’ gallery.}
    \label{fig:photoCollage}
\end{figure}

Given the particular topology of the site (fig. \ref{fig:photoCollage}), where the galleries are covered by a few meters wide flat field, it has been possible to record the changes in the muon flux crossing this piece of standard soil from Summer to Winter and study the correlations with the cumulative rain. Worth mentioning that the full project also foresees comparison of the muographic measurements with other geophysics measurements to assess the performance of this method with respect to more traditional surveys for shallow soil depths of the order of a few meters. The main result of the present study is that we observe an overall decrease with time of the recorded muon rates crossing the soil overburden and the increase of the total precipitation received by the soil during the same period. We propose as an explanation for this behavior the retention of water by the soil above the detector site. 

\section{Experimental setup}
\label{secExpSetup}
The detector we installed inside the gallery was a three scintillation-planes tracker. Particles that traverse all three planes are registered as a sequence of three ($x$,$y$,$z$) coordinates points where $z$ is the distance between planes, 63 cm between the extreme planes and 32.5 cm between the bottom detector and the middle one (fig. \ref{fig:photoCollage} – Bottom (right)). A linear regression fit is then applied for these points to determine whether they belong on the trace of a linear trajectory or not. Based on the chi-square value, along with other event selection criteria, we apply a cut to sample only tracks with high degree of linearity, since these are most likely to be muons. The dimensions of the planes (40 cm x 40 cm) define a maximum zenithal angle for the incoming muons, $\theta_{max}=42^\circ$.
% that correlates strongly with the azimuthal $\phi$ directions defined by the diagonals of the square planes. The maximum zenithal angle for the muon tracks that is allowed to have $\phi \in [0^{\circ},360^{\circ}]$, is $\theta_{max}^{iso}=32.5^\circ$, defined by the acmes of the extreme detector planes.

\begin{figure}[h]
\centering
\includegraphics[width=0.90\textwidth]{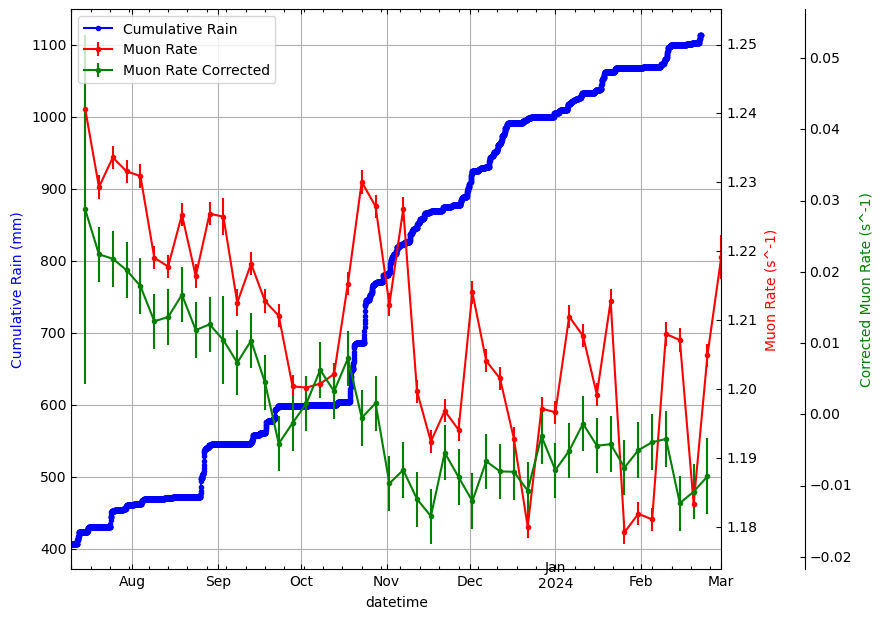}
\caption{Time series for the hourly cumulative precipitation (blue), the detected muon rates (red) and the residual muon rates (green) after accounting for the atmospheric pressure contribution (both aggregated over a period of 5 days).}
\label{fig:timeSeries}
\end{figure}

\begin{figure}[h]
\centering
    \begin{subfigure}[b]{0.45\textwidth}
    \centering
    \includegraphics[width=\textwidth]{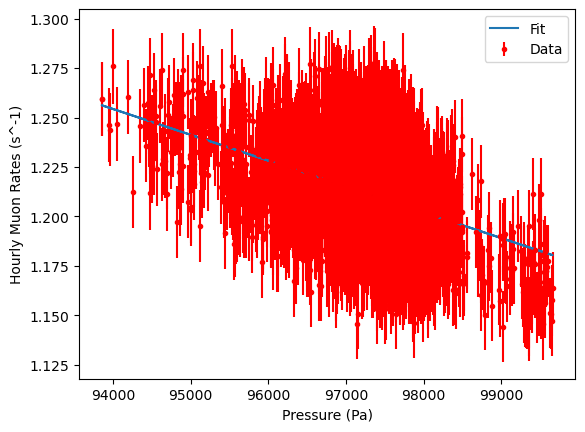}
    \caption{muon rates vs atmospheric pressure}
    \label{fig:muonRatesVsPresTotal}
    \end{subfigure}
    \hfill
    \begin{subfigure}[b]{0.45\textwidth}
    \centering
    \includegraphics[width=\textwidth]{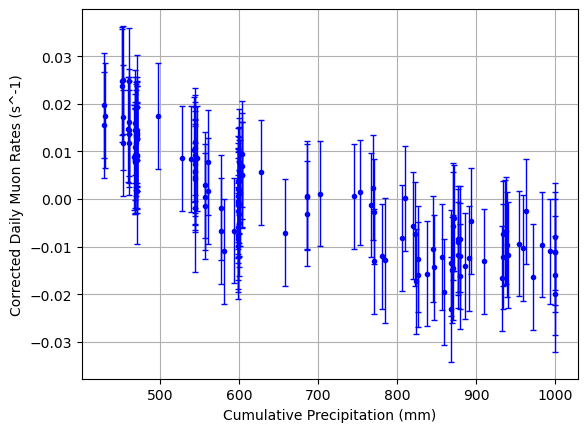}
    \caption{corrected daily muon rates vs cumulative precipitation}
    \label{fig:scatterPlot}
    \end{subfigure}
\caption{Scatter Plots: (a) Hourly muon rates vs atmospheric pressure. Linear Fit Results - Intercept~= ~$2.48 \pm 0.04\ s^{-1}$, Slope~=~$-131 \pm 4\times 10^{-7}s^{-1}Pa^{-1}$. (b) the corrected muon rates plotted against the cumulative precipitation shows clearly the correlation between the two variables (pearson: -0.84), the error bars give an indication of the large uncertainties for the relevant calculations in sections \ref{secSimulation} \& \ref{secSimComp}.}
\label{fig:relationCorrMuonsVscumPrec}
\end{figure}

We used environmental data taken from the Climate Data Store\footnote{https://cds.climate.copernicus.eu/}, specifically the ``Reanalysis'' atmospheric parameters dataset for the closest coordinate grid point to our site (N~$45.53^\circ$,~E~$4.86^\circ$) \citep{hersbach2017}. Weather data are provided in hourly values, and we aggregated the muon data with the same time step. It has been established in the past that there is a linear dependence of the muon rates to the local atmospheric pressure \citep{jourde2016}. To account for the impact of the pressure changes on the measured muon counts we fit the respective scatter plot (fig. \ref{fig:muonRatesVsPresTotal}) with a linear regression model, that we then use to correct the hourly muon rates. The mean value for the hourly rate during the DAQ was $r_0 = 1.20764\pm0.00025\ s^{-1}$ which corresponds to the $0.00 s^{-1}$ rate value for the corrected flux (green $y$ axis - fig. \ref{fig:timeSeries}). This information becomes relevant when one needs to calculate the relative change of the corrected muon rates.

Both the corrected daily muon rates (green points with error bars) and the experimental daily muon rate values (red points with error bars) show a decreasing trend which correlates with an increase of the cumulative precipitation (fig. \ref{fig:timeSeries}). We see that the muon rates from mid November 2023 until the end of February 2024 have stopped their downward trend and have stabilized. This could mean that the additional precipitation does not remain inside the soil overburden, and we can hypothesize that at this point the soil porosity is saturated with water, so that the additional water passes through. Under the assumption (see next paragraph) of 380 cm of soil that withholds a maximum precipitation of 800 mm we calculated a porosity of 21\%. The cumulative precipitation is measured from the start of the year and shows that by the initialization of our measurements the ground had already received $\sim$430 mm of water which at the end of the year had reached $\sim$1000 mm. This behavior is shown clearly when plotting the corrected daily muon rates as a function of the cumulative precipitation (fig. \ref{fig:scatterPlot}).

\subsection*{Overburden Opacity}
\label{subsecSoilHeightCalc}
The value for the soil height above the gallery shown in figure \ref{fig:photoCollage} is retrieved by a series of calculation based on the experimental data acquired during the first 96 days that returned an experimental muon rate value\footnote{The selection cuts used for the muon tagging at the time of this evaluation were stricter than the ones mentioned previously.} of $0.3326 \pm 0.0005 s^{-1}$. The overburden was hypothesized to comprise of standard rock (table 8 - \cite{groom2001}). The efficiency of the detector was evaluated based on measurements done at the last floor and the basement of the Dirac building, where IP2I of Lyon is operating. The materials of the building between the two positions amount to 3 m.w.e and with this prior we retrieved an overall efficiency for the detector equal to $0.1524 \pm 0.0004$. With this efficiency accounted for, the theoretical muon rate is $2.182 \pm 0.003 s^{-1}$ that corresponds to $360.7 \pm 0.6$ cm of standard rock overburden. Opacity is defined as the material density ($\rho$) times the length ($\ell$) of the muon trajectory within it. The overburden opacity for vertical muons is in this case 956 $g/cm^2$. In terms of opacity we subtract 400 mm of water and 16 cm of stone/cement mixture (sec. \ref{secSimulation}) for the gallery arch and we retrieve 380 cm of soil.

\section{Simulation}
\label{secSimulation}
The simulation we used is based on Geant4 libraries (\cite{geant4_agost2003}, \cite{geant4_allis2006}, \cite{geant4_allis2016}) and uses randomly generated muon tracks that follow the parametrisation described in \cite{shukla2018} for the particle energy and the zenithal track distribution while the azimuthal distribution is considered isotropic. The simulated detector follows the actual detector geometry. The sensitive material used for the scintillation parts is polyvinyltoluene (Luxium Solutions, BC-416) for the extreme panels and polyethylene for the middle one. The information returned from the simulation is the position, energy, energy deposition, momentum direction and the type of particle for each interaction with the detector's scintillation parts.

A valid muon event in this context is an event where the muon deposits energy on every detector volumes even for those that are accompanied by other particles (mainly electrons and gammas) which may also interact with the detector. In this sense the simulation will always tend to overestimate the signal since it doesn't take into account the detection efficiency and the event selection algorithm for the experimental data analysis.

The implementation of the different structures surrounding the detector is based on the visual inspection and measurements performed at the site but the materials used are arbitrarily selected since there is no sample catalog of the exact properties so much for the building blocks of the foundations as much for the composition of the soil above them. We assumed a mixture of cement (10\%) and rock (90\%) for the building foundations and a ``Clay Loam'' (33\% Clay, 33\% Silt, 34\% Sand) type of soil for the overburden. Silt and Sand for simplification purposes are considered to be both SiO\textsubscript{2} while the Clay is presumed to be pure Kaolinite ($Al_2Si_2O_5(OH)_4$), giving rise to a soil with 2.3 $g/cm^3$ density.

The porosity in this context is represented as a void volume, placed beneath the soil, which can be filled with different ratios of air and water. The water saturation of this volume at 100\% corresponds to 800 mm of water retention. In fig. \ref{fig:photoCollage} (bottom left) the water saturation drawn (blue strip) is 50\% (400 mm), that is similar to the value it should have at the initialization of the measurements. 

\section{Data-Monte Carlo comparison}
\label{secSimComp}
With the mean hourly muon rate (section \ref{secExpSetup}) as a reference value and by consulting fig. \ref{fig:scatterPlot} we calculate that the relative change of the muon rates from the start of the experiment until their stabilization in mid November is $2.3\% \pm 0.14\%$\footnote{The initial rate is calculated by the data collected the first three days as $r_0 = 1.2362\pm 0.0017 s^{-1}$ and the change in the muon rates until the stabilization is $\Delta r = 0.0286 \pm 0.0017 s^{-1}$, with the final corrected muon rate calculated for the entire period after mid November at $-0.0074 \pm 0.0005 s^{-1}$.}.

To study the response of the detector to the water retention effect we simulated ~3.8M muons that would be detected under open sky conditions, which in terms of experimental DAQ time corresponds to $\sim$10 days. We then run the simulation two times, the first for 50\% water saturation (or ~400mm cumulative precipitation) which resulted in 714816 detected muons. The second run was for 100\% water saturation (or ~800mm cumulative precipitation) for which 704269 muons were detected. The relative change for the detected muons between the two water configurations is 1.47\%.

This result is of the same order of magnitude with the experimental value which is encouraging given the simplicity of our hypotheses and the way that the environmental parameters have been acquired. It is clear that the experimental results and the simulation will be better constraint once the information from the geophysics surveys become available to us.

\section{Conclusions \& Outlook}
\label{conclusions}
In this note we present preliminary results on an hydrological survey performed with the muography technique, by placing a detector in an underground gallery topped with a few meters of soil. This study takes place in a more general study of an archaeological site, more precisely designed to characterize the shapes and depths of underground galleries not completely excavated. The site is located in Vienne, close to Lyon, France. 
The detector located underground recorded almost continuously atmospheric muons under a 3.8 meters overburden from June 2023 to February 2024. The changes in the measured atmospheric muon flux have been studied in correlation with the accumulated rain in the overburden soil. Given the very simple topology of the experimental setup, the number of parameters to be tuned in this analysis is relatively limited, which makes this configuration very powerful to further study the muography performance for hydrological surveys. 
Here we present a simplified model adjusting the porosity of the soil to reproduce the observed data. A Monte-Carlo simulation has been developed to allow for data-MC comparison and the model proves to be satisfactorily reproducing the data. The project will be continued in 2024 and the analysis improved by additional information from other geophysical surveys (ERT, georadar and distributed acoustic sensing) conducted recently and at present under analysis. The adjunction of a meteorological weather station would facilitate the method by giving more accurate insights for the local conditions, since the 0.5\% difference between the experiment and the simulation could be easily explained by the divergence of local precipitation to the values presented here.

\section*{Acknowledgements}
We would like to thank Laurence Brissaud, Jean-Luc Prisset for their initiative to include Arch\'emuons collaboration into their archaeological activities at Palais de Miroir, the direction of the Saint-Roman-en-Gal museum and specifically Giulia Ciucci for the support and the facilitation of the experimental activities. Finally we should acknowledge Christophe Benech (Archéorient) and Benoît Tauzin (LGL-TPE) contributions for setting up and performing the geophysics surveys.

This work has been carried out with the support of the LabEx LIO of the Université Claude Bernard Lyon 1, which has been created as part of a program for future development (reference number ANR-10-LABEX-066) initiated by the French government and overseen by the Agence Nationale de la Recherche (ANR). This study is also part of the ANR-19-CE05-0033 MEGAMU project.





\end{document}